\documentclass[aps,prl,twocolumn,showpacs,superscriptaddress,groupedaddress]{revtex4}
\usepackage{graphicx}  
\usepackage{dcolumn}   
\usepackage{bm}        	
\usepackage{amssymb}   
\usepackage{amsmath}
\usepackage{float}

\usepackage{natbib}

\usepackage{color}
\newcommand{\red}[1]{#1} 

\begin{document}

\title{Monte Carlo Sampling in Fractal Landscapes}

\author{Jorge C. Leit\~ao}
\email[Author to whom correspondence should be sent. E-mail address:
      ]{jleitao@pks.mpg.de}
\affiliation{Max Planck Institute for the Physics of Complex Systems, 01187 Dresden, Germany}
\affiliation{CFP and Faculdade de Ci\^{e}ncias da Universidade do Porto, 4169-007 Porto, Portugal}

\author{J. M. Viana Parente Lopes}
\affiliation{CEsA - Centre for Wind Energy and Atmospheric Flows and\\
Faculdade de Engenharia da Universidade do Porto, 4200-465 Porto, Portugal }

\author{Eduardo G. Altmann}
\affiliation{Max Planck Institute for the Physics of Complex Systems, 01187 Dresden, Germany}

\date{\today}

\begin{abstract}
We design a random walk to explore fractal landscapes such as those describing chaotic transients in dynamical systems. We show that the random walk moves efficiently only when its step length depends on the height of the landscape via the largest Lyapunov exponent of the system. We propose a generalization of the Wang-Landau algorithm which constructs not only the density of states (transient time distribution) but also the correct step length. As a result, we obtain a flat-histogram Monte Carlo method which samples fractal landscapes in polynomial time, a dramatic improvement over the exponential scaling of traditional uniform-sampling methods. Our results are not limited by the dimensionality of the landscape and are confirmed numerically in chaotic systems with up to 30 dimensions.
\end{abstract}

\maketitle

The development of Monte Carlo methods had a dramatic impact on our understanding of high-dimensional systems. The
spectrum of applications of these methods was considerably expanded with the development of optimized methods, such as
flat-histogram~\cite{Berg1991,Wang2001,Lopes2006} and parallel tempering~\cite{Swendsen1986} and now includes problems in a variety of fields, ranging from fluid dynamics~\cite{Yan2003} and spin systems~\cite{Berg1991,Wang2001,Lopes2006} to protein simulations~\cite{Trebst2006,Grassberger1997}.
These methods efficiently compute averages using nonuniform sampling and are optimized to problems on which the high dimensionality of the system leads to phase spaces with complex (rough) energy landscapes.

\begin{figure}[!ht]
\includegraphics[width=\linewidth]{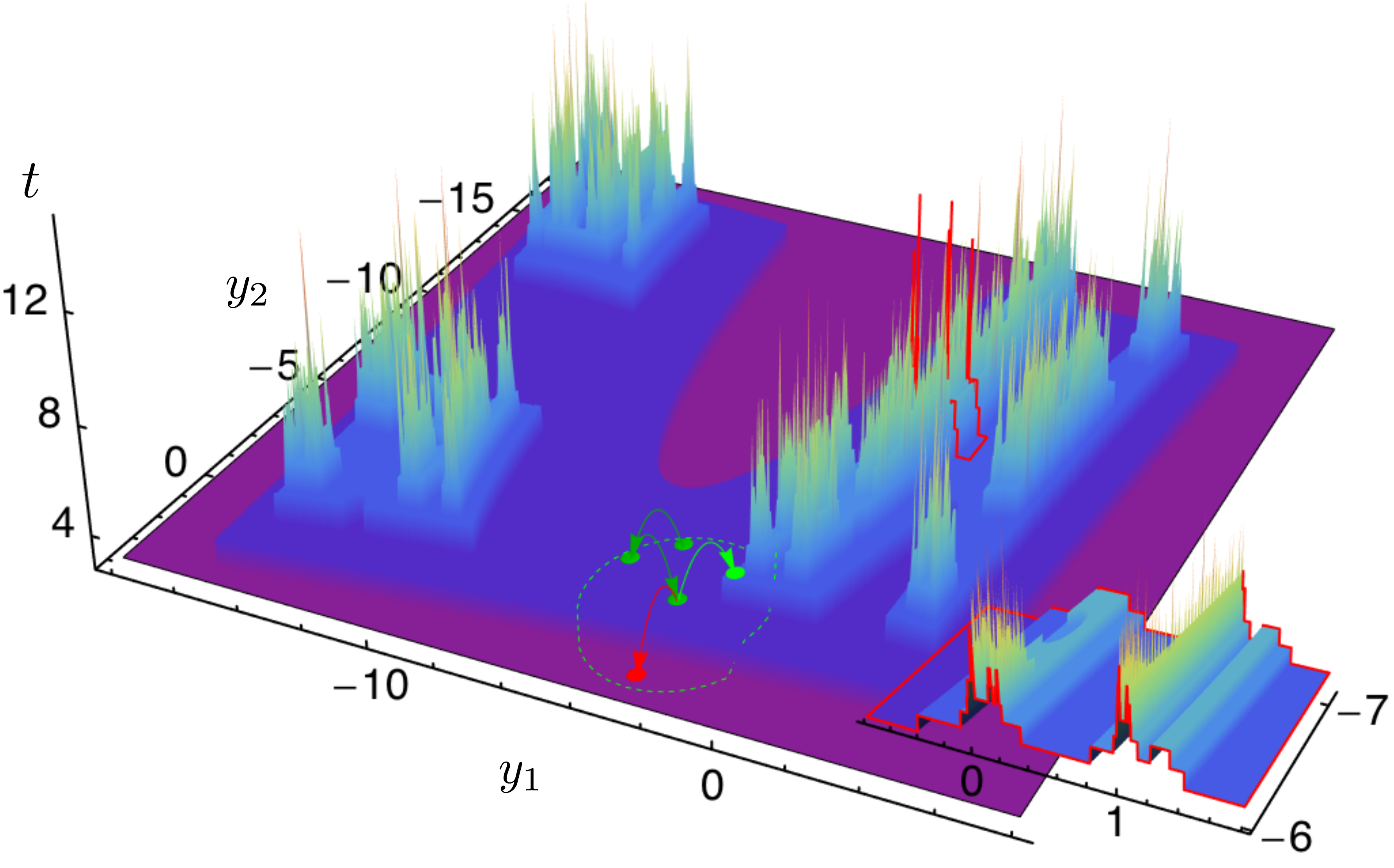}
\caption{
(Color online) Fractal landscapes in transient chaos.
Escape time $t$ as a function of the phase space coordinates $(y_1, y_2)$ at $x_1=x_2=0$ of the four-dimensional coupled H\'enon maps defined in Eq. (\ref{eq:maps})\red{, which will be given later}. Inset: magnification showing a Cantor set-like profile.
The circles (states) and arrows (proposals) represent the random walk (green/red indicate accepted/rejected proposals) underlying the Monte
Carlo sampling.
}
\label{fig1}
\end{figure}

In chaotic dynamical systems, complex landscapes appear even in low dimensions due to the sensitivity of initial conditions.
Prominent examples of such landscapes appear in systems showing chaotic transients. Transient chaos is a classical problem of nonlinear dynamics~\cite{OttBook} with recent applications in fields ranging from quantum scattering to chemical and biological reactions in fluid flows~\cite{LaiTamasBook, altmann.rmp}. 
\red{In transient chaotic systems, trajectories have a finite-time chaotic regime characterized by the time $t$ they need to escape the chaotic region of the phase-space.} The fraction $\rho(t)$ of initial conditions which escape the chaotic transient at time $t$ decays as $\rho(t) \sim e^{-\kappa t}$ (\red{where} $\kappa$ is the escape rate) and the set of initial conditions with $t=\infty$ is fractal \red{(e.g. a Cantor set)}~\cite{LaiTamasBook,Grassberger1997}. The dependence of $t$ on the phase-space \red{coordinates} $\boldsymbol{r}$ build thus a fractal landscape \red{where the escape time $t$ is interpreted as its height}, as illustrated in Fig.~\ref{fig1}.
Such extreme rough landscapes pose major numerical challenges~\cite{OttBook,LaiTamasBook}.
While algorithms beyond uniform sampling have been proposed for specific problems, e.g. to compute the fractal dimension~\cite{DeMoura2001} or to find long-living trajectories~\cite{Nusse1989,Sweet2001,Bollt2005}, there is still no general framework to sample the phase space of such systems.

In this Manuscript we show how Monte Carlo methods can be applied to fractal landscapes such as those appearing in dynamical systems with
chaotic transients.
The crucial step is to design a random walk able to sample the extreme roughness of fractal landscapes.
We show that an efficient flat-histogram simulation is only obtained using a random-walk step length~$\sigma$ which scales with the landscape height $t$ as $\sigma(t) \sim e^{-\lambda_L t}$, where $\lambda_L$ is the maximum Lyapunov exponent of the underlying chaotic system.
Moreover, by extending the Wang-Landau procedure~\cite{Wang2001} to the proposal distribution of random walk steps, we obtain an adaptive algorithm which provides simultaneously $\rho(t)$ and $\sigma(t)$.
In transient chaos problems, our approach changes the scaling of the computational effort from exponential to polynomial (with maximum $t$) and both efficiently finds the large $t$ trajectories and computes averages over the phase space.

We consider a fractal landscape as an escape time function of a transient chaotic system. Given a discrete-time open dynamical system $\boldsymbol{r}_{n+1} = \boldsymbol{F}(\boldsymbol{r}_n)$ defined in a $D$-dimensional phase space $\Omega$, the escape time $t(\boldsymbol{r})$ is defined as the number of iterations needed for an initial condition $\boldsymbol{r}$ to leave the region of nontrivial dynamics~\cite{LaiTamasBook}.
We propose an algorithm that constructs both the total volume $\rho(t)$ of the landscape \red{(which is the escape time distribution of the open chaotic system)} and the correct step length $\sigma(t)$ at each $t$, in a predetermined time spectrum $[t_\text{{min}},t_\text{{max}}]$ and with a precision f, which is successively reduced (initially $f=e$ and $\sigma(t)=\rho(t)=1$  for all $t$).
\red{The underlying random walk of the algorithm consists in: 1. proposing of a new state and 2. accepting or rejecting the proposed state. The random walk domain is the space} of initial conditions $\Gamma \in \Omega$~\cite{footnote1}, is initialized at $\boldsymbol{r} \in\Gamma, t = t(\boldsymbol{r})$, and evolves according to the following four steps:

\begin{enumerate}
	\item[{\bf S1-}] propose a state $\boldsymbol{r}' \in \Gamma$ with
$t' = t(\boldsymbol{r}')\in[t_\text{{min}}, t_\text{{max}}]$ [e.g.,
            using Eq.~(\ref{eq:proposal}) below]. 
	\item[{\bf S2-}] accept/reject the state $\boldsymbol{r}'$ according to flat-histogram choice [Eq.~(\ref{eq:acceptance}) below].
	\item[{\bf S3-}] update $\rho(t)$ and $\sigma(t)$, respectively, to:
\begin{itemize}
\item[{\bf S3.1-}] $\rho(t) f$ (Wang-Landau); 
\item[{\bf S3.2-}] $\sigma(t) f$ if $t' = t$; $\sigma(t)/f$ if $t' < t$.
\end{itemize}
	\item[{\bf S4-}] After a number of repetitions of {\bf S1}-{\bf S3}, refine $f$ to $\sqrt{f}$ and go to {\bf S1}.

\end{enumerate}
This procedure stops when $f = f_\text{min} \gtrsim 1$, a value which controls the precision of $\rho(t)$ and $\sigma(t)$. Using only {\bf S1} and {\bf S2}, the random-walk corresponds to a flat-histogram Monte Carlo simulation on $t$~\cite{Berg1991}. We now describe in more detail the steps {\bf S1-S4}, \red{see Supplementary Material for an implementation of the method.}

\noindent{\bf{S1-Proposal}} - The ideal random walk should be able to explore the order of the landscape for an efficient search. In discrete spaces, often considered in spin systems, there is a natural local step given by flipping a single spin~\cite{NewmanBarkemaBook}. In continuous spaces the locality of the step is determined by the neighborhood around the present state.
Fractal landscapes do not have a global characteristic length scale~\cite{OttBook,LaiTamasBook} and therefore we consider a height dependent step length $\sigma=\sigma(t)$. Accordingly, we choose an isotropic conditional probability of proposing a new state $\boldsymbol{r}'$ given $\boldsymbol{r}$ as
\begin{equation}
g(\boldsymbol{r}\rightarrow\boldsymbol{r}') = \frac{1}{\sigma(t(\boldsymbol{r}))}e^{-|\boldsymbol{r} - \boldsymbol{r}'|/\sigma(t(\boldsymbol{r}))}\ \ ,
\label{eq:proposal}
\end{equation}
where $\sigma(t)$ gives the characteristic length of the distribution~\cite{footnote2}.

\begin{figure}[!ht]
\includegraphics[width=\linewidth]{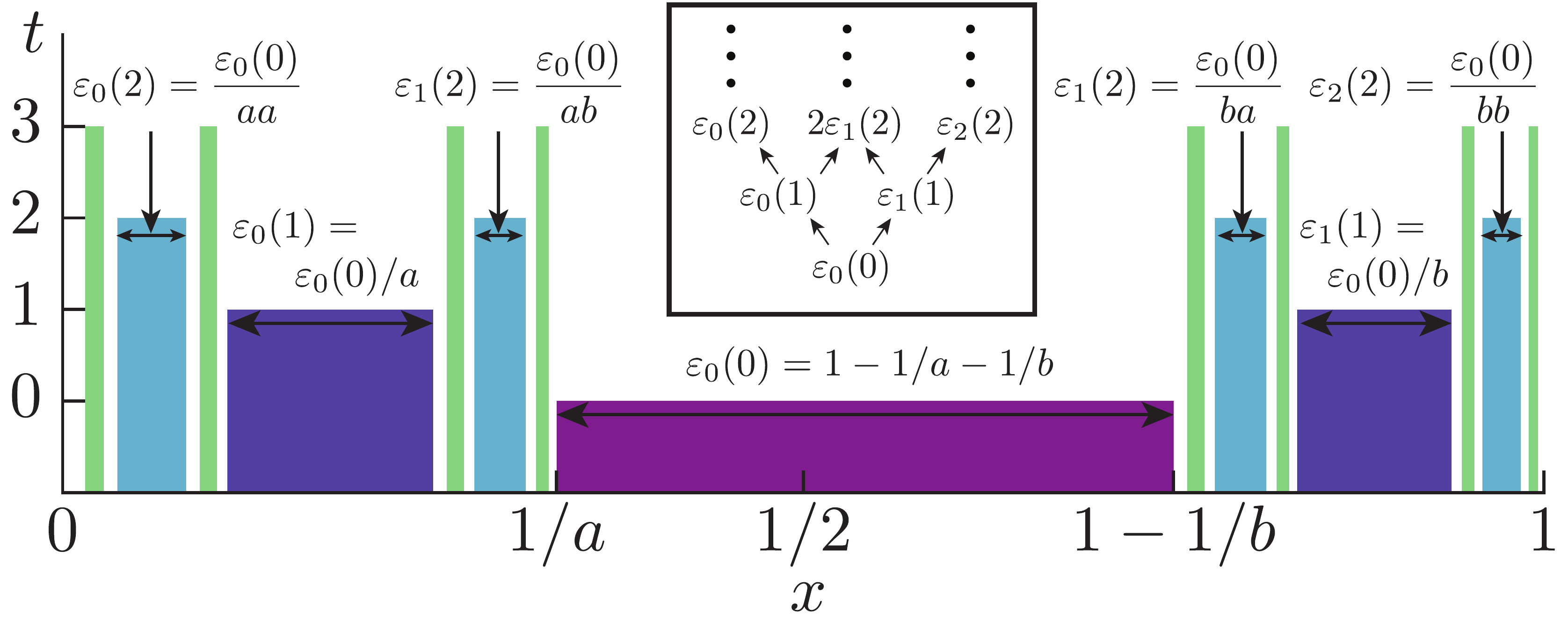}
\caption{
(Color online) \red{Fractal landscape corresponding to the $2$-scale Cantor set with scales $1/a$ and $1/b$}. A plateau at $t$ with width $\varepsilon(t)$ generates two plateaus at $t+1$, with widths $\varepsilon(t+1)=\varepsilon(t)/a$ and  $\varepsilon(t+1)=\varepsilon(t)/b$, see inset.
At each $t$, the $2^t$ plateaus have lengths $\varepsilon_k(t)$ with $k=0 \ldots t$.
}
\label{fig1Prime}
\end{figure}

We now show how $\sigma(t)$ has to scale with $t$ for an efficient proposal. We consider the construction of the Cantor set~\cite{OttBook,LaiTamasBook} as a paradigm of fractal landscape \red{appearing in transient chaotic systems}, see Fig. \ref{fig1Prime}. The construction starts by splitting the interval $[0,1]$ in the intervals $[0,1/a],\ [1/a,1-1/b],\ [1-1/b,1]$ and assigning the escape time $t=0$ to the middle interval (plateau at $t=0$). This procedure is repeated on each of the two surviving intervals by assigning $t=1$ to each of their two middle intervals (plateaus at $t=1$), and again in the remaining intervals \emph{ad infinitum}. 
In order to achieve an efficient proposal we have to know the scaling of the typical length of the plateaus $\tilde{\varepsilon}(t)$ with $t$.
For the one-scale Cantor set ($a = b$), each of the $2^t$ plateaus have a unique length given by $\varepsilon(t) = (1 - 2/a)(1/a)^t$ and thus $\tilde{\varepsilon}(t) = \varepsilon(t)$.
For the two-scale Cantor set ($a \neq b$), the $2^t$ plateaus have $t+1$ different lengths $\varepsilon_k(t) = (1 - 1/a - 1/b) (1/a)^{t-k}(1/b)^k$ with
$k=0,...,t$ and the number of plateaus with size $\varepsilon_k(t)$ is the binomial coefficient $B(t,k)$, see inset of Fig.~\ref{fig1Prime}. The total
length at $t$ is $\rho(t) = (1 - 1/a - 1/b)(1/a + 1/b)^t \sim \exp(-\kappa t)$. The conditional probability of being at a plateau of length $\varepsilon_k(t)$ at a given $t$ is
\begin{equation}\label{eq.conditional}
P(k|t) =\frac{P(k,t)}{P(t)} = \frac{B(t,k) \varepsilon_k(t)}{\rho(t)}\ \ .
\end{equation}
The characteristic plateau size is thus naturally chosen as
$\tilde{\varepsilon}(t) = \varepsilon_{k^*}(t)$ where $k=k^*$ maximizes $P(k|t)$ in Eq.~(\ref{eq.conditional}). Using Stirling's approximation we obtain $k^* \approx t/(1+b/a)$ and thus
\begin{equation}
\tilde{\varepsilon}(t) = \varepsilon_{k^*}(t) = \exp(-t \frac{a\log(b) + b\log(a)}{a+b})\ \ .
\label{eq:epsilon}
\end{equation}
In the context of transient chaos, the construction of the Cantor set corresponds exactly to the escape time function of the
one-dimensional open tent map~\cite{footnote3}, and the exponent $\lambda_L=\frac{a\log(b) + b\log(a)}{a+b}$ corresponds to its positive Lyapunov
exponent~\cite{LaiTamasBook}.
This leads to the following natural interpretation for a choice of $\sigma(t)$ with $\tilde{\varepsilon}(t)$ as given in Eq.~(\ref{eq:epsilon}): in order to ensure that two chaotic trajectories (initiated at~$r$ and $r'$) remain correlated up to time $t$, their initial distance
$|r-r'|$ should be reduced exponentially with $t$, with an exponent equal to the positive Lyapunov exponent responsible for the divergence
in forward time. In a generic fractal landscape, generated by a higher-dimensional system, this divergence is dominated by the maximal Lyapunov exponent~$\lambda_L$ and therefore 
\begin{equation}
\sigma(t) \sim \tilde{\varepsilon}(t) \sim e^{-\lambda_L t}\ \ 
\label{eq:sigma}
\end{equation}
should be used in any isotropic proposal such as Eq.~(\ref{eq:proposal}).

\begin{figure}[!ht]
\includegraphics[width=\linewidth]{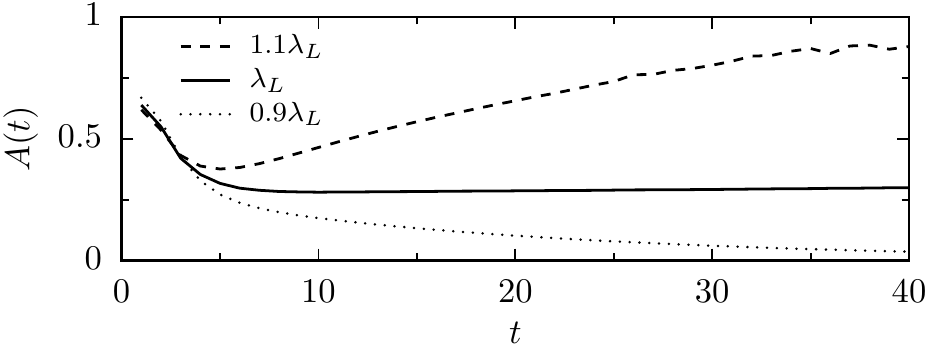}
\caption{
Characteristic random-walk step~$\sigma$ has to scale as the typical plateau size~$\tilde{\varepsilon}$ in order to achieve a constant
acceptance ratio in time. Acceptance ratio~$A(t)$ of a flat-histogram simulation on the two-scale Cantor set with $(a,b)=(3,4)$
with three (see legend) different exponents $\lambda$ on the step length $\sigma(t)\sim e^{-\lambda t}$, with $\lambda_L$ given after Eq.~(\ref{eq:epsilon}).
For $\lambda < \lambda_L$ and $t\gg1$, $A(t)$ decays (exponentially) as a consequence of lack of proposals to $t'>t$ because $\sigma(t) \gg \tilde{\varepsilon}(t)$.
For $\lambda > \lambda_L$ and $t\gg1$, $A(t)$ increases to 1 but the simulation gets stuck in the same plateau as all proposals are for $t'=t$ because $\sigma(t) \ll \tilde{\varepsilon}(t)$.}
\label{fig2}
\end{figure}

\noindent{\bf{S2 - Acceptance}} - Because of the extreme roughness of fractal landscapes, we use a flat-histogram simulation~\cite{Berg1991} on the variable $t$, which plays the role traditionally played by energy. \red{In a flat-histogram, the probability to sample a state $\boldsymbol{r}$ is $1/\rho(t(\boldsymbol{r}))$.} Consequently, the detailed balance of this Monte Carlo process is fulfilled when the conditional probability of accepting a proposed state $\boldsymbol{r}'$ given $\boldsymbol{r}$ follows the Metropolis's choice~\cite{NewmanBarkemaBook}
\begin{equation}
A(\boldsymbol{r}\rightarrow\boldsymbol{r}') = \min \left\{1, \frac{\rho(t(\boldsymbol{r}))}{\rho(t(\boldsymbol{r}'))} \frac{g(\boldsymbol{r}'\rightarrow\boldsymbol{r})}{g(\boldsymbol{r}\rightarrow\boldsymbol{r}')}  \right\}\ \ ,
\label{eq:acceptance}
\end{equation}
where $g(\boldsymbol{r}\rightarrow\boldsymbol{r}')$ is given by Eq.~(\ref{eq:proposal}). Since we are considering projections in $t$, it is useful to define the conditional probability $A(t)$ of accepting a proposal given a time $t$~\cite{NewmanBarkemaBook}. In the spirit of flat-histogram simulations, a signature of an efficient random walk is an $A(t)$ which does not strongly depends on $t$.
In Fig.~\ref{fig2} we show that only when the scaling in Eq.~(\ref{eq:sigma}) is used in the Eq.~(\ref{eq:proposal}), we obtain a constant $A(t)$ and thus an efficient simulation.

\noindent{\bf{S3 - Wang-Landau update}} - In systems on which $\rho(t)$ and $\sigma(t)$ (or $\lambda_L$) are known, we use steps {\bf S1-S2} to sample them. However, for generic landscapes, $\rho(t)$ and $\sigma(t)$ are unknown. We take advantage of the analogy between $\rho(t)$ and a density of states and apply the Wang-Landau procedure to compute it~\cite{Wang2001}. This is done by successive approximating $\rho(t)$ in steps {\bf S3} and {\bf S4} of our approach.
To compute $\sigma(t)$, we propose the following generalization of the Wang-Landau procedure (step {\bf S3.1}) to the proposal distribution (step {\bf S3.2}): if the proposed state has an escape time smaller than the present state, $t' < t$, we decrease $\sigma(t)$ by dividing it by $f$. If it has the same escape time, $t' = t$, we increase $\sigma(t)$ by multiplying it by $f$. Asymptotically ($f = f_\text{min} \rightarrow 1$), a flat-histogram Markov process is recovered.

\noindent{\bf{S4 - Refinement}} - Steps {\bf S1}-{\bf S3} are repeated for a predefined number of round-trips~\cite{NewmanBarkemaBook, Costa2007}, defined as the movement in the time-spectrum from $t_\text{{min}}$ to $t_\text{{max}}$ and back to $t_\text{{min}}$.  The number of round-trips is chosen using an equivalent procedure to the one in Ref.~\cite{Belardinelli2007}. After that, we refine the precision parameter~$f$ by taking its square root~\cite{Wang2001}).

\begin{figure}[H]
\includegraphics[width=\linewidth]{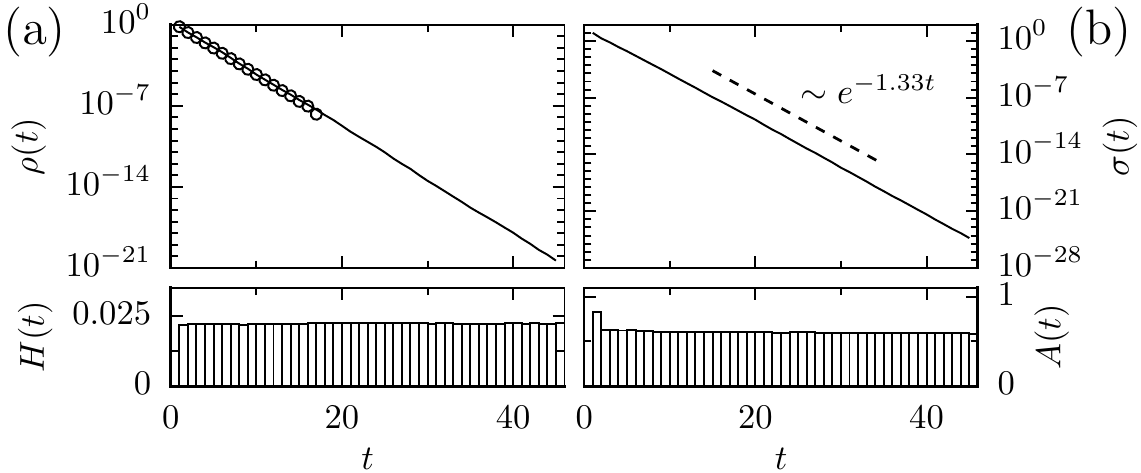}
\caption{Confirmation that our method yields the correct values of $\kappa$ and $\lambda_L$ and converges to a flat
  histogram simulation for the case $N=2$ in Eq.~(\ref{eq:maps}). (a) $\rho(t)$ obtained through our method (line) and through uniform
  sampling (circles) with the same computational effort \red{(measured in number of map iterations)}. Lower inset: histogram $H(t)$ of visits to escape times $t$.  (b) $\sigma(t)$
  obtained through our method. The dashed line shows the scaling $e^{-\lambda_L t}$ with $\lambda_L\approx1.33$ obtained in
  Ref.~\cite{Sweet2001}.  Lower inset: the acceptance
  ratio $A(t)$. We used $[t_\text{{min}},t_\text{{max}}]=[1,45]$, $\log_e f_{min} = 2^{-13}$ and all quantities were measured on the last
  refinement.
}
\label{fig3}
\end{figure}

We now confirm the generality of the approach described above through numerical simulations in generic fractal landscapes generated by a family of coupled H\'enon maps $\boldsymbol{r}_{n+1}=F(\boldsymbol{r}_n)$, with $\boldsymbol{r}=\{x_i,y_i\}_{i=1}^N$  and $F$ defined by
\begin{equation}
\left(\begin{array}{c}
x_{i}\\
y_{i}
\end{array}\right)_{n+1} = 
\left(\begin{array}{c}
A_i - x_i^2 + B y_i + k(x _i - x_{i+1})\\
x_i
\end{array}\right)_n,
\label{eq:maps}
\end{equation}
with $i=1,...,N$, $N + 1 \equiv 1$ and parameters $k = 0.4$, $B = 0.3$, $A_1 = 3$ (if $N>1$), $A_N = 5$, and $A_{i} = A_1 + (A_N -
A_1)(i-1)/(N-1)$.  This choice of parameters ensures that a chaotic map is obtained in the $N=1$ case and the map considered in
Ref.~\cite{Sweet2001} is recovered for $N=2$ (used as a representative case to illustrate our algorithm). Initial conditions are on a $2N$
hypercube $\Gamma=[-4,4]^{2N}$ and escape is defined as leaving $\Gamma$. In Fig.~\ref{fig3} we confirm the convergence and validity of our algorithm by showing that the computed $\rho(t)$ coincides with the one obtained using uniform sampling, $\sigma(t)$ scales with the Lyapunov exponent reported in Ref.~\cite{Sweet2001}, and both the acceptance and the histogram of visits to escape time $t$ are flat in $t$.

We now compare our approach to uniform sampling in terms of computational efficiency. For each $t_\text{{max}}$, we compute the average number
of map iterations $n(t_\text{max})$ per sampled state with $t=t_\text{max}$. This comparison guarantees that the uncertainty of any observable at $t = t_\text{max}$ (worst case) is the same in both approaches.
For a uniform-sampling simulation, $n(t_\text{max}) \sim 1/\rho(t_\text{max}) \sim e^{\kappa t_\text{max}}$. For a flat-histogram simulation, obtained after the convergence of our method {\bf S1}-{\bf S4}, we adopt a conservative approach which avoids the sampling of correlated states by considering a single sample of $t_\text{max}$ for each round-trip.
The estimation of $n(t_\text{max})$ in this case is based on the expected number of steps per round-trip expected of an unbiased random walk in the time spectrum with local steps ($\Delta t \approx 1$), which scales as $\sim t_\text{{max}}^2$. Additionally, each proposal requires $t$ map iterations and, since the histogram is flat, for each round trip one gets an additional $t_\text{max}$ contribution, leading to an expected scaling of $n\sim t_\text{max}^3$.
\begin{figure}[!ht]
\includegraphics[width=\linewidth]{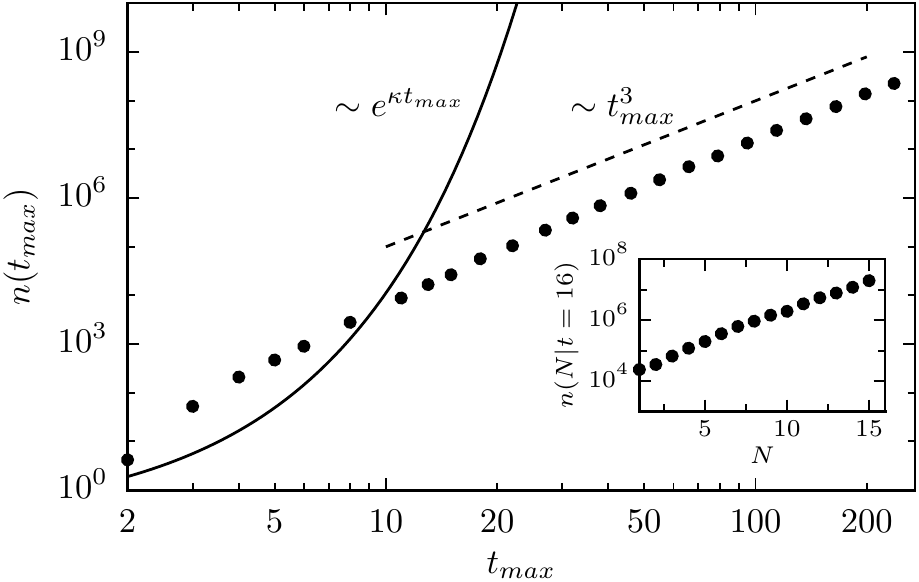}
\caption{
The computation effort in number of map iterations $n(t_\text{max})$ scales polynomially with maximum escape time, in contrast to the
exponential scaling of uniform sampling. Main panel: results for uniform sampling (solid line) and flat-histogram simulation achieved by our
method (squares) for the representative case $N=4$ in Eq.~(\ref{eq:maps}). The dashed line indicates the scaling $t_\text{max}^3$. Inset: dependence of the efficiency on the phase-space dimension $2N$ at a fixed $t_{\text{max}} = 16$.
}
\label{fig4}
\end{figure}
Figure~\ref{fig4} confirms the dramatic improvement from exponential (uniform sampling) to polynomial (our approach) scaling in the coupled
H\'enon maps.
The significance of these results become apparent by noticing that $t_\text{max}=237$ (last point in Fig.~\ref{fig4}) corresponds to
$\rho(t_\text{max})\approx10^{-109}$, meaning that we are able to sample extremely rare states. For such level of accuracy, our
method requires an implementation with arbitrary precision~\cite{GMP} which in our case was able to resolve states which differ by $10^{-137}$
[since $\sigma(t_\text{max}=237) \approx 10^{-137}$].
Interestingly, the slight but clear deviation from the prediction $t_\text{max}^3$ seen in Fig.~\ref{fig4} shows that flat-histogram
simulations on fractal landscapes are not purely diffusive on $t$, a phenomenon known in spin-systems as critical slowing down~\cite{Dayal2004,Trebst2004}. This phenomenon is enhanced with increasing dimension and contributes to the exponential increase of $n_\text{max}$ with $N$ for a fixed $t_\text{max}$, as shown in the inset of Fig.~\ref{fig4}. Still, an uniform sampling in such a high-dimension ($2N=30$) phase-space would
need impracticable $n\approx 10^{34}$ map iterations to sample one state with $t_\text{max}=16$.

In summary, we have shown how flat-histogram Monte Carlo simulations can be performed on fractal landscapes.
The crucial ingredient is to consider a random-walk step size dependent on the height of the landscape. The correct 
dependency should scale as the characteristic length of the landscape and can be obtained through an adaptive procedure
which generalizes Wang-Landau's algorithm to the proposal distribution. This idea can find applications in any rough landscape with a height dependent characteristic width.
Fractality can be considered as an extreme case of roughness which naturally occurs in dynamical systems with chaotic transients. In this case, our results show that the Lyapunov exponent $\lambda_L$, a fundamental property of the chaotic dynamics, is an essential ingredient for a flat-histogram simulation.

We emphasize the significance of our results for numerical investigations of transient chaos. Our method automatically provides
\red{the escape rate $\kappa$ and the maximum Lyapunov exponent of the system $\lambda_L$}, is not limited to low dimension, and allows for the computation of expected values of any observable using a flat-histogram simulation. For the specific problem of finding the chaotic saddle~\cite{Nusse1989,Sweet2001,Bollt2005}, which is indirectly solved in our simulations by storing trajectories with large $t$, our findings show that best results are achieved using a proposal which scales as $e^{-\lambda_L t}$.

More generally, besides high dimensionality, the sensitivity of initial conditions in chaotic systems is a major reason for using
statistical methods in physics.
Monte Carlos methods in dynamical systems were traditionally limited to uniform sampling and, only recently, optimized methods (with
nonuniform sampling) were applied for the problem of finding trajectories with low chaoticity~\cite{Yanagita2009,Tailleur2007}.
Our approach opens the perspective of using the full strength of optimized Monte Carlo methods in problems that involve the computation
of averages in chaotic systems. Spatially extended~\cite{Tel2008} and nonhyperbolic Hamiltonian~\cite{Cristadoro2008} systems are natural candidates for future applications of this approach.

We are indebted to T. T\'el and P. Grassberger for insightful discussions. J.C.L. acknowledges funding from Erasmus Grant No. 29233-IC-1-2007-1-PT-ERASMUS-EUCX-1 and Max Planck Society.


\begin{thebibliography}{24}%
\makeatletter
\providecommand \@ifxundefined [1]{%
 \@ifx{#1\undefined}
}%
\providecommand \@ifnum [1]{%
 \ifnum #1\expandafter \@firstoftwo
 \else \expandafter \@secondoftwo
 \fi
}%
\providecommand \@ifx [1]{%
 \ifx #1\expandafter \@firstoftwo
 \else \expandafter \@secondoftwo
 \fi
}%
\providecommand \natexlab [1]{#1}%
\providecommand \enquote  [1]{``#1''}%
\providecommand \bibnamefont  [1]{#1}%
\providecommand \bibfnamefont [1]{#1}%
\providecommand \citenamefont [1]{#1}%
\providecommand \href@noop [0]{\@secondoftwo}%
\providecommand \href [0]{\begingroup \@sanitize@url \@href}%
\providecommand \@href[1]{\@@startlink{#1}\@@href}%
\providecommand \@@href[1]{\endgroup#1\@@endlink}%
\providecommand \@sanitize@url [0]{\catcode `\\12\catcode `\$12\catcode
  `\&12\catcode `\#12\catcode `\^12\catcode `\_12\catcode `\%12\relax}%
\providecommand \@@startlink[1]{}%
\providecommand \@@endlink[0]{}%
\providecommand \url  [0]{\begingroup\@sanitize@url \@url }%
\providecommand \@url [1]{\endgroup\@href {#1}{\urlprefix }}%
\providecommand \urlprefix  [0]{URL }%
\providecommand \Eprint [0]{\href }%
\providecommand \doibase [0]{http://dx.doi.org/}%
\providecommand \selectlanguage [0]{\@gobble}%
\providecommand \bibinfo  [0]{\@secondoftwo}%
\providecommand \bibfield  [0]{\@secondoftwo}%
\providecommand \translation [1]{[#1]}%
\providecommand \BibitemOpen [0]{}%
\providecommand \bibitemStop [0]{}%
\providecommand \bibitemNoStop [0]{.\EOS\space}%
\providecommand \EOS [0]{\spacefactor3000\relax}%
\providecommand \BibitemShut  [1]{\csname bibitem#1\endcsname}%
\let\auto@bib@innerbib\@empty
\bibitem [{\citenamefont {Berg}\ and\ \citenamefont
  {Neuhaus}(1991)}]{Berg1991}%
  \BibitemOpen
  \bibfield  {author} {\bibinfo {author} {\bibfnamefont {B.~A.}\ \bibnamefont
  {Berg}}\ and\ \bibinfo {author} {\bibfnamefont {T.}~\bibnamefont {Neuhaus}},\
  }\href@noop {} {\bibfield  {journal} {\bibinfo  {journal} {Phys. Lett. B}\
  }\textbf {\bibinfo {volume} {267}},\ \bibinfo {pages} {249} (\bibinfo {year}
  {1991})}\BibitemShut {NoStop}%
\bibitem [{\citenamefont {Wang}\ and\ \citenamefont {Landau}(2001)}]{Wang2001}%
  \BibitemOpen
  \bibfield  {author} {\bibinfo {author} {\bibfnamefont {F.}~\bibnamefont
  {Wang}}\ and\ \bibinfo {author} {\bibfnamefont {D.P.}~\bibnamefont {Landau}},\
  }\href@noop {} {\bibfield  {journal} {\bibinfo  {journal} {Phys. Rev. Lett.}\
  }\textbf {\bibinfo {volume} {86}},\ \bibinfo {pages} {2050} (\bibinfo {year}
  {2001})}\BibitemShut {NoStop}%
\bibitem [{\citenamefont {{Viana Lopes}}\ \emph {et~al.}(2006)\citenamefont
  {{Viana Lopes}}, \citenamefont {Costa}, \citenamefont {Lopes dos Santos},\ and\
  \citenamefont {Toral}}]{Lopes2006}%
  \BibitemOpen
  \bibfield  {author} {\bibinfo {author} {\bibfnamefont {J.}~\bibnamefont
  {{Viana Lopes}}}, \bibinfo {author} {\bibfnamefont {M.D.}~\bibnamefont
  {Costa}}, \bibinfo {author} {\bibfnamefont {J.M.B.}~\bibnamefont {Lopes dos Santos}}, \
  and\ \bibinfo {author} {\bibfnamefont {R.}~\bibnamefont {Toral}},\
  }\href@noop {} {\bibfield  {journal} {\bibinfo  {journal} {Phys. Rev. E}\
  }\textbf {\bibinfo {volume} {74}},\ \bibinfo {pages} {046702} (\bibinfo
  {year} {2006})}\BibitemShut {NoStop}%
\bibitem [{\citenamefont {Swendsen}\ and\ \citenamefont
  {Wang}(1986)}]{Swendsen1986}%
  \BibitemOpen
  \bibfield  {author} {\bibinfo {author} {\bibfnamefont {R.~H.}\ \bibnamefont
  {Swendsen}}\ and\ \bibinfo {author} {\bibfnamefont {J.-S.}\ \bibnamefont
  {Wang}},\ }\href@noop {} {\bibfield  {journal} {\bibinfo  {journal} {Phys.
  Rev. Lett.}\ }\textbf {\bibinfo {volume} {57}},\ \bibinfo {pages} {2607}
  (\bibinfo {year} {1986})}\BibitemShut {NoStop}%
\bibitem [{\citenamefont {Yan}\ and\ \citenamefont {de~Pablo}(2003)}]{Yan2003}%
  \BibitemOpen
  \bibfield  {author} {\bibinfo {author} {\bibfnamefont {Q.}~\bibnamefont
  {Yan}}\ and\ \bibinfo {author} {\bibfnamefont {J.J.}~\bibnamefont {de~Pablo}},\
  }\href@noop {} {\bibfield  {journal} {\bibinfo  {journal} {Phys. Rev. Lett.}\
  }\textbf {\bibinfo {volume} {90}},\ \bibinfo {pages} {035701} (\bibinfo
  {year} {2003})}\BibitemShut {NoStop}%
\bibitem [{\citenamefont {Trebst}\ \emph {et~al.}(2006)\citenamefont {Trebst},
  \citenamefont {Troyer},\ and\ \citenamefont {Hansmann}}]{Trebst2006}%
  \BibitemOpen
  \bibfield  {author} {\bibinfo {author} {\bibfnamefont {S.}~\bibnamefont
  {Trebst}}, \bibinfo {author} {\bibfnamefont {M.}~\bibnamefont {Troyer}}, \
  and\ \bibinfo {author} {\bibfnamefont {U.~H.~E.}\ \bibnamefont {Hansmann}},\
  }\href@noop {} {\bibfield  {journal} {\bibinfo  {journal} {J. Chem. Phys.}\
  }\textbf {\bibinfo {volume} {124}},\ \bibinfo {pages} {174903} (\bibinfo
  {year} {2006})}\BibitemShut {NoStop}%
\bibitem [{\citenamefont {Grassberger}(1997)}]{Grassberger1997}%
  \BibitemOpen
  \bibfield  {author} {\bibinfo {author} {\bibfnamefont {P.}~\bibnamefont
  {Grassberger}},\ }\href@noop {} {\bibfield  {journal} {\bibinfo  {journal}
  {Phys. Rev. E}\ }\textbf {\bibinfo {volume} {56}},\ \bibinfo {pages} {3682}
  (\bibinfo {year} {1997})}\BibitemShut {NoStop}%
\bibitem [{\citenamefont {Ott}(1993)}]{OttBook}%
  \BibitemOpen
  \bibfield  {author} {\bibinfo {author} {\bibfnamefont {E.}~\bibnamefont
  {Ott}},\ }\href@noop {} {\emph {\bibinfo {title} {{Chaos in Dynamical
  Systems}}}} (\bibinfo  {publisher}
  {Cambridge University Press},\ \bibinfo {address} {Cambridge},\ \bibinfo
  {year} {1993}),\ \bibinfo {edition} {2nd}\ ed\BibitemShut {NoStop}%
\bibitem [{\citenamefont {Lai}\ and\ \citenamefont
  {T\'{e}l}(2011)}]{LaiTamasBook}%
  \BibitemOpen
  \bibfield  {author} {\bibinfo {author} {\bibfnamefont {Y.-C.}\ \bibnamefont
  {Lai}}\ and\ \bibinfo {author} {\bibfnamefont {T.}~\bibnamefont {T\'{e}l}},\
  }\href@noop {} {\emph {\bibinfo {title} {{Transient Chaos: Complex Dynamics
  in Finite Time Scales}}}},\ \bibinfo {series} {Applied Mathematical
  Sciences}\ (\bibinfo  {publisher} {Springer},\
  \bibinfo {year} {2011}), Vol.\ \bibinfo {volume} {173}\BibitemShut {NoStop}%
\bibitem [{\citenamefont {Altmann}\ \emph {et~al.}(2013)\citenamefont
  {Altmann}, \citenamefont {Portela},\ and\ \citenamefont
  {T\'{e}l}}]{altmann.rmp}%
  \BibitemOpen
  \bibfield  {author} {\bibinfo {author} {\bibfnamefont {E.~G.}\ \bibnamefont
  {Altmann}}, \bibinfo {author} {\bibfnamefont {J.~S.~E.}\ \bibnamefont
  {Portela}}, \ and\ \bibinfo {author} {\bibfnamefont {T.}~\bibnamefont
  {T\'{e}l}},\ }\href@noop {} {\bibfield  {journal} {\bibinfo  {journal} {Rev. Mod. Phys. 85, 869Ð918}\ } (\bibinfo {year} {2013})}\BibitemShut {NoStop}%
\bibitem [{\citenamefont {de~Moura}\ and\ \citenamefont
  {Grebogi}(2001)}]{DeMoura2001}%
  \BibitemOpen
  \bibfield  {author} {\bibinfo {author} {\bibfnamefont {A.~P.~S.}\
  \bibnamefont {de~Moura}}\ and\ \bibinfo {author} {\bibfnamefont
  {C.}~\bibnamefont {Grebogi}},\ }\href@noop {} {\bibfield  {journal} {\bibinfo
   {journal} {Phys. Rev. Lett.}\ }\textbf {\bibinfo {volume} {86}},\ \bibinfo
  {pages} {2778} (\bibinfo {year} {2001})}\BibitemShut {NoStop}%
\bibitem [{\citenamefont {Nusse}\ and\ \citenamefont
  {Yorke}(1989)}]{Nusse1989}%
  \BibitemOpen
  \bibfield  {author} {\bibinfo {author} {\bibfnamefont {H.~E.}\ \bibnamefont
  {Nusse}}\ and\ \bibinfo {author} {\bibfnamefont {J.~A.}\ \bibnamefont
  {Yorke}},\ }\href@noop {} {\bibfield  {journal} {\bibinfo  {journal} {Physica
  D}\ }\textbf {\bibinfo {volume} {36}},\ \bibinfo {pages} {137} (\bibinfo
  {year} {1989})}\BibitemShut {NoStop}%
\bibitem [{\citenamefont {Sweet}\ \emph {et~al.}(2001)\citenamefont {Sweet},
  \citenamefont {Nusse},\ and\ \citenamefont {Yorke}}]{Sweet2001}%
  \BibitemOpen
  \bibfield  {author} {\bibinfo {author} {\bibfnamefont {D.}~\bibnamefont
  {Sweet}}, \bibinfo {author} {\bibfnamefont {H.~E.}\ \bibnamefont {Nusse}}, \
  and\ \bibinfo {author} {\bibfnamefont {J.~A.}\ \bibnamefont {Yorke}},\
  }\href@noop {} {\bibfield  {journal} {\bibinfo  {journal} {Phys. Rev. Lett.}\
  }\textbf {\bibinfo {volume} {86}},\ \bibinfo {pages} {2261} (\bibinfo {year}
  {2001})}\BibitemShut {NoStop}%
\bibitem [{\citenamefont {Bollt}(2005)}]{Bollt2005}%
  \BibitemOpen
  \bibfield  {author} {\bibinfo {author} {\bibfnamefont {E.~M.}\ \bibnamefont
  {Bollt}},\ }\href@noop {} {\bibfield  {journal} {\bibinfo  {journal} {Int. J.
  Bifurcat. Chaos}\ }\textbf {\bibinfo {volume} {15}},\ \bibinfo {pages} {1615}
  (\bibinfo {year} {2005})}\BibitemShut {NoStop}%
\bibitem [{\citenamefont {footnote1}(footnote1)}]{footnote1}%
  \BibitemOpen
  \bibfield  {author} {\bibinfo {author} {Which intersects the stable manifold of the chaotic saddle}}\BibitemShut {NoStop}%
\bibitem [{\citenamefont {footnote2}(footnote2)}]{footnote2}%
  \BibitemOpen
  \bibfield  {author} {\bibinfo {author} {We verified that a normal distribution with standard deviation $\sigma(t)$ gives equivalent results}}\BibitemShut {NoStop}%
\bibitem [{\citenamefont {footnote3}(footnote3)}]{footnote3}%
  \BibitemOpen
  \bibfield  {author} {\bibinfo {author} {The tent map is defined on $x\in[0,1]$ as $x_{t+1} = a x_t$ for $x_t < b/(a+b)$ and $x_{t+1} = b (1 - x_t)$ for $x_t > b/(a+b)$~\cite{LaiTamasBook}}}\BibitemShut {NoStop}%
\bibitem [{\citenamefont {Newman}\ and\ \citenamefont
  {Barkema}(2002)}]{NewmanBarkemaBook}%
  \BibitemOpen
  \bibfield  {author} {\bibinfo {author} {\bibfnamefont {M.~E.~J.}\
  \bibnamefont {Newman}}\ and\ \bibinfo {author} {\bibfnamefont {G.~T.}\
  \bibnamefont {Barkema}},\ }\href@noop {} {\emph {\bibinfo {title} {{Monte
  Carlo Methods in Statistical Physics}}}}\ (\bibinfo  {publisher} {Oxford
  University},\ \bibinfo {address} {New York},\ \bibinfo {year}
  {2002})\BibitemShut {NoStop}%
\bibitem [{\citenamefont {Costa}\ \emph {et~al.}(2005)\citenamefont {Costa},
  \citenamefont {{Viana Lopes}},\ and\ \citenamefont {dos Santos}}]{Costa2007}%
  \BibitemOpen
  \bibfield  {author} {\bibinfo {author} {\bibfnamefont {M.~D.}\ \bibnamefont
  {Costa}}, \bibinfo {author} {\bibfnamefont {J.}~\bibnamefont {{Viana
  Lopes}}}, \ and\ \bibinfo {author} {\bibfnamefont {J.~M. B.~L.}\ \bibnamefont
  {dos Santos}},\ }\href@noop {} {\bibfield  {journal} {\bibinfo  {journal}
  {Europhys. Lett.}\ }\textbf {\bibinfo {volume} {72}},\ \bibinfo {pages} {802}
  (\bibinfo {year} {2007})}\BibitemShut {NoStop}%
\bibitem [{\citenamefont {Belardinelli}\ and\ \citenamefont
  {Pereyra}(2007)}]{Belardinelli2007}%
  \BibitemOpen
  \bibfield  {author} {\bibinfo {author} {\bibfnamefont {R.E.}~\bibnamefont
  {Belardinelli}}\ and\ \bibinfo {author} {\bibfnamefont {V.D.}~\bibnamefont
  {Pereyra}},\ }\href@noop {} {\bibfield  {journal} {\bibinfo  {journal} {Phys.
  Rev. E}\ }\textbf {\bibinfo {volume} {75}},\ \bibinfo {pages} {046701}
  (\bibinfo {year} {2007})}\BibitemShut {NoStop}%
\bibitem [{\citenamefont {Granlund}\ and\ \citenamefont {the GMP
  Development~Team}(2012)}]{GMP}%
  \BibitemOpen
  \bibfield  {author} {\bibinfo {author} {\bibfnamefont {T.}~\bibnamefont
  {Granlund}}\ and\ \bibinfo {author} {\bibnamefont {the GMP
  Development~Team}},\ }\href@noop {} {\enquote {\bibinfo {title} {{GNU MP}},}\
  } (\bibinfo {year} {2012})\BibitemShut {NoStop}%
\bibitem [{\citenamefont {Dayal}\ \emph {et~al.}(2004)\citenamefont {Dayal},
  \citenamefont {Trebst}, \citenamefont {Wessel}, \citenamefont {Wurtz},
  \citenamefont {Troyer}, \citenamefont {Sabhapandit},\ and\ \citenamefont
  {Coppersmith}}]{Dayal2004}%
  \BibitemOpen
  \bibfield  {author} {\bibinfo {author} {\bibfnamefont {P.}~\bibnamefont
  {Dayal}}, \bibinfo {author} {\bibfnamefont {S.}~\bibnamefont {Trebst}},
  \bibinfo {author} {\bibfnamefont {S.}~\bibnamefont {Wessel}}, \bibinfo
  {author} {\bibfnamefont {D.}~\bibnamefont {Wurtz}}, \bibinfo {author}
  {\bibfnamefont {M.}~\bibnamefont {Troyer}}, \bibinfo {author} {\bibfnamefont
  {S.}~\bibnamefont {Sabhapandit}}, \ and\ \bibinfo {author} {\bibfnamefont
  {S.~N.}\ \bibnamefont {Coppersmith}},\ }\href@noop {} {\bibfield  {journal}
  {\bibinfo  {journal} {Phys. Rev. Lett.}\ }\textbf {\bibinfo {volume} {92}},\
  \bibinfo {pages} {097201} (\bibinfo {year} {2004})}\BibitemShut {NoStop}%
\bibitem [{\citenamefont {Trebst}\ \emph {et~al.}(2004)\citenamefont {Trebst},
  \citenamefont {Huse},\ and\ \citenamefont {Troyer}}]{Trebst2004}%
  \BibitemOpen
  \bibfield  {author} {\bibinfo {author} {\bibfnamefont {S.}~\bibnamefont
  {Trebst}}, \bibinfo {author} {\bibfnamefont {D.~A.}\ \bibnamefont {Huse}}, \
  and\ \bibinfo {author} {\bibfnamefont {M.}~\bibnamefont {Troyer}},\
  }\href@noop {} {\bibfield  {journal} {\bibinfo  {journal} {Phys. Rev. E}\
  }\textbf {\bibinfo {volume} {70}},\ \bibinfo {pages} {046701} (\bibinfo
  {year} {2004})}\BibitemShut {NoStop}%
\bibitem [{\citenamefont {Yanagita}\ and\ \citenamefont
  {Iba}(2009)}]{Yanagita2009}%
  \BibitemOpen
  \bibfield  {author} {\bibinfo {author} {\bibfnamefont {T.}~\bibnamefont
  {Yanagita}}\ and\ \bibinfo {author} {\bibfnamefont {Y.}~\bibnamefont {Iba}},\
  }\href@noop {} {\bibfield  {journal} {\bibinfo  {journal} {J. Stat.
  Mech.-Theory E}\ ,\ \bibinfo {pages} {P02043}} (\bibinfo {year}
  {2009})}\BibitemShut {NoStop}%
\bibitem [{\citenamefont {Tailleur}\ and\ \citenamefont
  {Kurchan}(2007)}]{Tailleur2007}%
  \BibitemOpen
  \bibfield  {author} {\bibinfo {author} {\bibfnamefont {J.}~\bibnamefont
  {Tailleur}}\ and\ \bibinfo {author} {\bibfnamefont {J.}~\bibnamefont
  {Kurchan}},\ }\href@noop {} {\bibfield  {journal} {\bibinfo  {journal} {Nat.
  Phys.}\ }\textbf {\bibinfo {volume} {3}},\ \bibinfo {pages} {203} (\bibinfo
  {year} {2007})}\BibitemShut {NoStop}%
\bibitem [{\citenamefont {T\'{e}l}\ and\ \citenamefont {Lai}(2008)}]{Tel2008}%
  \BibitemOpen
  \bibfield  {author} {\bibinfo {author} {\bibfnamefont {T.}~\bibnamefont
  {T\'{e}l}}\ and\ \bibinfo {author} {\bibfnamefont {Y.-C.}\ \bibnamefont
  {Lai}},\ }\href@noop {} {\bibfield  {journal} {\bibinfo  {journal} {Phys.
  Rep.}\ }\textbf {\bibinfo {volume} {460}},\ \bibinfo {pages} {245} (\bibinfo
  {year} {2008})}\BibitemShut {NoStop}%
\bibitem [{\citenamefont {Cristadoro}\ and\ \citenamefont
  {Ketzmerick}(2008)}]{Cristadoro2008}%
  \BibitemOpen
  \bibfield  {author} {\bibinfo {author} {\bibfnamefont {G.}~\bibnamefont
  {Cristadoro}}\ and\ \bibinfo {author} {\bibfnamefont {R.}~\bibnamefont
  {Ketzmerick}},\ }\href@noop {} {\bibfield  {journal} {\bibinfo  {journal}
  {Phys. Rev. Lett.}\ }\textbf {\bibinfo {volume} {100}},\ \bibinfo {pages}
  {184101} (\bibinfo {year} {2008})}\BibitemShut {NoStop}%
\end{thebibliography}
\end{document}